\documentclass{article}
\usepackage[utf8]{inputenc}
\usepackage{amsmath}
\DeclareMathOperator{\Tr}{Tr}
\title{On the Peierls Interpretation of Quantum Mechanics}
\author{M G Burt, Department of Physics, Durham University, Durham, DH1 3LE, UK}
\begin{document}
\maketitle
\begin{abstract}
The brief works of Peierls on the role of the observer in quantum mechanics are examined, interpreted and expanded to widen accessibility and understanding of these works. The approach followed here is very much in the spirit adopted by Peierls who eschewed a `rigorous axiomatic' and aimed at using `logic at the level of the working physicist'. The fundamental tenet of his work is that the wavefunction or density matrix represents the knowledge of an observer and that two observers of the same system may well have different knowledge and will use different density matrices to describe it. Essential to the understanding of Peierls's approach is the demonstration given here that the density matrix generally can be expressed entirely in terms of probabilities of observable outcomes and that such probabilities are subjective in the first instance.The process by which the knowledge of two observers may be combined is discussed in detail for some simple cases and, by using Bayes's theorem, the progress to a common understanding of the correct density matrix for a given situation demonstrated. Peierls gave a criterion to ensure that the amalgamation of data from two observers can never lead to violation of the uncertainty principle : that the density matrices of two observers must commute. It is pointed out that it is essential that the density matrices of two observers commute under all circumstances no matter how inaccurate or even fanciful the data or beliefs used to construct them. Once this is appreciated it can be shown that Peierls's density matrix commutation criterion is fully equivalent to the standard result in quantum mechanics that two observables must commute if they are to be known precisely and simultaneously, but now extended to imperfectly known mixed states rather than a single eigenstate. Peierls's second criterion for two observers to have compatible density matrices, that their product is non zero, is discussed and clarified.
\end{abstract}

\section{Introduction}
The theorist Rudolf Peierls obtained his doctorate under Heisenberg and was subsequently Pauli's  research assistant for 3 years$^1$. His interpretation of quantum mechanics is based on Heisenberg's idea that the wavefunction, or more generally the density matrix, is a representation of one's knowledge about the quantum system being considered. By introducing an observer or observers into the theory, the interpretation affords an explanation for the collapse or contraction of the wavefunction, an explanation accessible to the nonspecialist. His exposition is further attractive to such a nonspecialist because it is close to what one may loosely call the conventional and familiar Copenhagen Interpretation, and, by highlighting the role of the observer, avoids the measurement paradox$^2$ that plagues the version$^3$ in which the observer plays no role.

Peierls put forward his interpretation in a short article in Physics World$^4$ as a reply to Bell's criticism$^5$ of the foundations of quantum theory in the same journal; the article essentially summarized the content of  a number of general interest book chapters$^{6,7,8}$. These expositions are very readable since Peierls eschewed `a rigorous axiomatic' in favour of an exposition `at the level of the logic of the working physicist'. However, these expositions were remarkably short and there are some points which need expansion to satisfy the inquiring reader, an expansion the present author aims to provide in the spirit of Peierls's expositions i.e. a rendition aimed at a user of quantum mechanics rather than a specialist in its foundations. In particular, Peierls put forward$^{4,8}$ a requirement for two observers of a system, that their density matrices must commute to avoid any possible violations to the uncertainty principle. This is an important requirement to understand since it touches the essence of the scientific method that observers should share knowledge and reach a common understanding of the physical system studied.

The essential feature of the Peierls interpretation, as already indicated, is that the wavefunction or more generally the density matrix expresses the knowledge of an observer. The knowledge may well be incomplete and even wildly inaccurate; he mentions that an observer `may only have an incomplete or inaccurate view of the measuring instrument'$^8$. Of course, although Peierls does not discuss this, it is obvious that this can only be describing the first step in the scientific process. With any disparity between the observers' knowledge, more experiments are performed until a consensus is reached among the scientific community as to the correct density  matrix to be assigned to a given system. This process is carried out, if not explicitly, then implicitly, using Bayes's theorem to modify the probability of a hypothesis being correct in the light of data collected. The question immediately arises as to how this process can be applied to quantum theory which famously involves not only probabilities, but also probability amplitudes; indeed, the off diagonal elements of the density matrix are often referred to as containing phase information which has a probability amplitude `ring' to it. However, it will be shown ( section 2 ) how to express the elements of any density matrix of finite dimension entirely in terms of the expectation values of observables of the system, which in turn can be expressed entirely in terms of the probabilities of the various possible outcomes of measurement of those observables. All information about probability amplitudes, so crucial to understanding quantum phenomena in general, is contained in a density matrix so constructed. Such probabilities as appear in the density matrix can be modified in the light of new data using Bayes's theorem and with sufficient high quality observations a consensus can be achieved as to the correct values to be assigned to them, an example of which is given in section 3.
	
It is within this general framework that we  examine ( section 4 ) Peierls's requirement that the density matrices of two observers of the same system must commute. It is shown that this requirement is equivalent to the familiar one that  the observables used by the two observers must commute if their measurements are never to interfere with each other, but now extended to imperfectly known mixed states rather than a single eigenstate. It will be seen that, no matter how inaccurate the data supplied by the two observers, if the commutation criterion is always satisfied, their combined density matrix will always have non-negative eigenvalues and hence can never violate the uncertainty principle as will be shown explicitly. On the other hand, we will see that if Peierls's requirement is not met, then, with sufficiently inaccurate data ( inaccuracy which is not due to the uncertainty principle ) , violations of the uncertainty principle, i.e. density matrices with one or more negative eigenvalues, are possible. Of course, with sufficiently accurate data it is impossible to violate the uncertainty principle since no physically  or mathematically realizable state can do so.

In section 4 we will also clarify Peierls's other criterion that the product of the density matrices of two observers must not be zero in order to avoid contradiction. We show generally that if the Peierls product criterion is not obeyed, i.e. the product of the observers' density matrices does in fact equal zero, then no amount of further experimental information and the use of Bayes's theorem will remedy the situation. In section 5 we will re-examine some of Peierls's statements in the light of our work and demonstrate how we can clarify and augment them.

Throughout this article the points will be illustrated using a concrete example, the Stern-Gerlach experiment, which focuses on the magnetic moment due to the outer electron of the silver atom. In the first instance, we assume that two observers, in separate labs, set up identical sources of silver atoms, investigate the polarization properties of their respective beams and then compare results. But we will also consider the case of two observers making sequential measurements on the same source. And to illustrate the theoretical concepts discussed we will imagine that we have an apparatus capable of measuring the nuclear magnetic moment of the atoms as well. Although it is far fetched in practice, it is possible in principle and will be a very useful illustrative aid.

\section{The Density Matrix expressed in terms of Probabilities}
 We consider the two dimensional, i.e. spin $1/2$ , case here leaving the general case of higher dimensional state spaces to Appendices A and B. We will show that all the elements of the density matrix , $\rho$ , can be written in terms of probabilities of the outcome of measurements of observables, in this simple case the $x$, $y$ and $z$ components of the spin. Consider first , $\langle \sigma_z \rangle$ , the quantum mechanical expectation value ( or rather twice that value ) of the $z$ component of the spin, represented by the appropriate Pauli matrix. We have
 \begin{equation}
 \langle\sigma_z\rangle = \Tr[\rho \sigma_z]=\rho_{1,1}-\rho_{2,2}
 \end{equation}
 But we also have from the normalization of  $\rho$ 
  \begin{equation}
\Tr[\rho]=\rho_{1,1}+\rho_{2,2}=1
 \end{equation}
 and so
  \begin{equation}
\rho_{1,1} =( 1 + \langle \sigma_z  \rangle)/2
 \end{equation}
 and
   \begin{equation}
\rho_{2,2} =( 1 - \langle \sigma_z  \rangle)/2
 \end{equation}
 Turning our attention to the $x$ component of the spin
  \begin{equation}
 \langle \sigma_x \rangle= \Tr[\rho \sigma_x]=\rho_{1,2}+\rho_{2,1}
 \end{equation}
 and finally for the $y$ component,
  \begin{equation}
 i\langle \sigma_y \rangle= \Tr[\rho i\sigma_y]=-\rho_{1,2}+\rho_{2,1}
 \end{equation}
 so that
  \begin{equation}
 \rho_{1,2} =(\langle \sigma_x \rangle- i \langle \sigma_y \rangle)/2
 \end{equation}
 and
 \begin{equation}
 \rho_{2,1} =(\langle \sigma_x \rangle+ i \langle \sigma_y \rangle)/2
 \end{equation}
 the complex conjugate as it must be. The above results ( equations (3) (4) (7) and (8) ) can be combined into the single matrix equation 
 
 \begin{equation}
\rho=(\frac{1}{2})(I + \langle \sigma_x \rangle \sigma_x +\langle \sigma_y \rangle \sigma_y + \langle \sigma_z \rangle \sigma_z )
\end{equation}
and its extension to all finite dimensions is given in Appendix A .

Now, to establish that these results for the elements of $\rho$ can be expressed in terms of probabilities of results of measurements, one notes that each $ \sigma_\alpha $ , where $\alpha =(x,y,z)$ , has eigenvalues $\pm1$ and hence
 \begin{equation}
 \langle \sigma_\alpha \rangle=p^{(\alpha,+1)}-p^{(\alpha,-1)}
 \end{equation}
 where $p^{(\alpha,+1)}$ and $p^{(\alpha,-1)}$ are the probabilities for measuring eigenvalues $+1$ and $-1$ respectively for $\sigma_\alpha $. From $p^{(\alpha,+1)}+p^{(\alpha,-1)}=1$ we readily find that $\rho_{1,1}$ and $\rho_{2,2}$ have particularly simple forms 
\begin{equation}
\rho_{1,1} = p^{(z,+1)}
\end{equation}
and
\begin{equation}
\rho_{2,2} = p^{(z,-1)}
\end{equation}
while $ \rho_{2,1} $ and  $\rho_{1,2}$ can be written as
\begin{equation}
 \rho_{2,1} = ((p^{(x,+1)}-p^{(x,-1)})+i(p^{(y,+1)}-p^{(y,-1)}))/2
\end{equation}
and
\begin{equation}
 \rho_{1,2} = ((p^{(x,+1)}-p^{(x,-1)})-i(p^{(y,+1)}-p^{(y,-1)}))/2
\end{equation}

So we have shown that the density matrix for a spin $1/2$ system can be expressed entirely in terms of the probabilities of the results of measurements. Even the off diagonal elements that provide the phase relationship between spin up and spin down states for the $z$ direction are determined by the probabilities of measurement  results, those  for the spin along the $x$ and $y$ directions.  As such, all the elements of  the density matrix can be updated as new information/measurement results become available using Bayes's theorem.
We now readily see why complete ignorance is represented essentially by the unit matrix$^4$. In the absence of any information, by The Principle of Indifference $^{9}$ we must have $p^{(\alpha,+1)}=p^{(\alpha,-1)}=1/2$ for each and every $\alpha$ giving
\begin{equation}
\rho = (1/2) \begin{pmatrix}
1 &0\\
0 &1
\end{pmatrix}
\end{equation}
Also, if we only have results for the measurement of $\sigma_z$ , then the density matrix must be diagonal.
 The result that the density matrix for a spin $1/2$ system can be expressed entirely in terms of the probabilities of the results of measurements can be extended to any number of finite dimensions as shown in Appendix B.

 \section{Constructing a Density Matrix from Data}
 Often the density matrix for a system is discussed as  a purely theoretical construct. But we need to appreciate that it can be derived directly from suitable data. We will now do this for the simplest of cases, the Stern-Gerlach experiment with an inhomogeneous field in the $z$ direction. We will assume the beam is sufficiently weak that the spin $1/2$ atoms pass through one at a time so that the detectors can act as counters and we record data $D(n_\uparrow,n_\downarrow)$ where $n_\uparrow$and $n_\downarrow$ are the number of counts corresponding to spin up and spin down respectively. From our considerations in the previous section we know that the density matrix is determined by $\langle \sigma_z  \rangle$, the quantum mechanical expectation value, but , of course, the data do not determine $\langle \sigma_z  \rangle$ precisely and the best we can do is compute a probability density distribution for the possible values, $\sigma$, of $\langle \sigma_z  \rangle$ . We can do this using Bayes's theorem. Normally one would consider no prior knowledge or prejudice and by The Principle of Indifference$^{9}$ use a constant , $1/2$ , for the prior probability distribution since $-1\leq\sigma\leq1$ and all values are equally likely. But it will be useful to consider the general case and assume a nonuniform  prior probability density distribution, $p_0(\sigma)$.Using Bayes's theorem,
\begin{equation}
p(\sigma\mid D)=\frac{p(D\mid \sigma) p_0(\sigma)}{p(D)}
\end{equation}
where $p(\sigma\mid D)$ is the probability density distribution for $\langle \sigma_z  \rangle$ given the data, $D$, $p(D\mid \sigma)$ is the probability of obtaining the data, $D$ , given $\langle \sigma_z  \rangle$ has the value $\sigma$, and $p(D)$ is the probability of obtaining data , $D$ , regardless of the value of $\langle \sigma_z  \rangle$, which is given by
\begin{equation}
p(D)= \int_{-1} ^{+1} p(D\mid \sigma) p_0(\sigma) \,\mathrm{d} \sigma
\end{equation}
and which, further, ensures that $ \int_{-1} ^{+1} p(\sigma\mid D)\,\mathrm{d} \sigma =1$ , as it must.

If the expectation value of $ \sigma_z $ is $\sigma$ , then the probability for spin up/down for  a single count is $(1 \pm \sigma)/2$ and $p(D\mid \sigma)$ is given by the binomial distribution :
\begin{equation}
 p(D\mid \sigma)=\binom{n}{n_\uparrow}(\frac{1+\sigma}{2})^{n_\uparrow}(\frac{1-\sigma}{2})^{n_\downarrow}
\end{equation}
where $n = n_\uparrow + n_\downarrow $ . This gives us our formula
\begin{equation}
p(\sigma\mid D)= \frac{(1+\sigma)^{ n_\uparrow }(1-\sigma)^{ n_\downarrow }p_0(\sigma)}{\int_{-1} ^{+1} (1+\sigma)^{ n_\uparrow }(1-\sigma)^{ n_\downarrow } p_0(\sigma) \,\mathrm{d} \sigma}
\end{equation}
for updating our knowledge in the light of data obtained. In particular, this formula illustrates the epitome of the scientific process when we examine the limit of large number of data points. In that limit $(1+\sigma)^{ n_\uparrow }(1-\sigma)^{ n_\downarrow }$ becomes sharply peaked at $\sigma_p=( n_\uparrow - n_\downarrow)/ (n_\uparrow + n_\downarrow)$ and $p(\sigma\mid D)$ becomes independent of $p_0(\sigma)$. This is an example of the more general principle, that sufficient data makes previous prejudices or errors irrelevant, which is discussed in Appendix C.
To obtain our effective density matrix as the result of collecting data, $D$ , we need to consider  $\langle O \rangle_D$ the expectation value of a general observable, $O$ ,  using our current probability distribution , $p(\sigma\mid D)$, for $\sigma$. If the value of $\langle \sigma_z  \rangle$  is taken as $\sigma$, then the expectation value of $O$ will be given by
\begin{equation}
\langle O \rangle_\sigma = \Tr[O \rho_\sigma ]
\end{equation}
where
\begin{equation}
\rho_\sigma = (1/2)\begin{pmatrix}
1+\sigma & 0 \\
0 & 1-\sigma
\end{pmatrix}
\end{equation}
But all values of $\sigma$ are possible with probability density distribution $p(\sigma\mid D)$. So the expectation value, $\langle O \rangle_D$ , of $O$ given data $D$,  is given by 
\begin{equation}
\langle O \rangle_D = \int_{-1} ^{+1} p(\sigma\mid D) \langle O \rangle_\sigma \,\mathrm{d} \sigma = \int_{-1} ^{+1} p(\sigma\mid D) \Tr[O \rho_\sigma ] \,\mathrm{d} \sigma
\end{equation}
or
\begin{equation}
\langle O \rangle_D  = \Tr[\int_{-1} ^{+1} p(\sigma\mid D) O \rho_\sigma  \,\mathrm{d} \sigma]=\Tr[O \rho_D ]
\end{equation}
where
\begin{equation}
\rho_D = \int_{-1} ^{+1} p(\sigma\mid D)\rho_\sigma \,\mathrm{d} \sigma
\end{equation}
 Assuming we started from complete ignorance, so that, $p_0(\sigma)$ ,  is just constant ( and equal to 1/2 for all $\sigma$ ), then it is straightforward to derive the effective density matrix 
 \begin{equation}
 \rho_D = \frac{1}{(n+2)}\begin{pmatrix}
n_{\uparrow}+1 & 0 \\
0 & n_{\downarrow}+1
\end{pmatrix}
 \end{equation}
 As a check , one notes that for no data at all, $n_{\uparrow}=n_{\downarrow}=0$ , we just obtain $(14)$ representing complete ignorance, while in the limit of large data sets we obtain probabilities $n_{\uparrow}/n$ and $n_{\downarrow}/n$ respectively for spin up and spin down.
 The reader will also be able to verify the intuitively reasonable result that, if an observer collects some data , $D^{(1)}=(n_{\uparrow}^{(1)},n_{\downarrow}^{(1)})$ , and calculates a probability distribution, $p(\sigma\mid D^{(1)})$ , and then a second observer takes this as prior information and updates it using their own data, $D^{(2)}=(n_{\uparrow}^{(2)},n_{\downarrow}^{(2)})$ , to obtain a probability distribution , $p(\sigma\mid (D^{(1)} \emph{ then } D^{(2)}))$ ,  then this will be the same as $p(\sigma\mid D^{(total)})$ where  $D^{(total)}=(n_{\uparrow}^{(1)}  + n_{\uparrow}^{(2)},n_{\downarrow}^{(1)}+n_{\downarrow}^{(2)})$, that is, the probability distribution obtained if one just amalgamated the two data sets. This is the simplest case of two observers collaborating, simplest because they are both measuring the same observable. We now move on to the case of two or more observers measuring different observables and examine the meaning and significance of the Peierls Commutation and Product criteria.

 \section{The Peierls Commutation and Product Criteria}
 Peierls$^4$ asserts that the density matrices of different observers must commute, otherwise one may violate the uncertainty principle. It is not difficult to come up with what appears to be a simple counter example. Suppose two observers , $Z$ and $X$ have identical Stern-Gerlach apparata including identical sources of silver atoms. They each measure the number of spin up and spin down atoms along the $z$ and $x$ directions respectively. They summarize their results by each producing their own density matrices, $\rho^{(Z)}$ and $\rho^{(X)}$ given by
 \begin{equation}
 \rho^{(Z)} = (\frac{1}{2})(I + \langle \sigma_z \rangle \sigma_z)
 \end{equation}
 and
 \begin{equation}
 \rho^{(X)} = (\frac{1}{2})(I + \langle \sigma_x \rangle \sigma_x)
 \end{equation}
 where $Z$ and $X$  respectively calculate $\langle \sigma_z \rangle$ and $\langle \sigma_x \rangle$ from their own data. Observer $Z$ must set $\langle \sigma_x \rangle=\langle \sigma_y \rangle=0$ in (9), by The Principle of Indifference $^9$ ,  as they have no knowledge of the polarization in the $x$ and $y$ directions. Similarly $X$ must set $\langle \sigma_y \rangle=\langle \sigma_z \rangle=0$ . The commutator of the observers'  density matrices is readily found to be
\begin{equation}
[\rho^{(Z)},\rho^{(X)}]=(i/4)\langle \sigma_z \rangle \langle \sigma_x \rangle \sigma_y \neq 0
\end{equation}
This commutator will only vanish if at least one of the expectation values vanishes  which is not the case in general. So we seem to have a contradiction here. Peierls's commutation criterion seems to be saying that $Z$ and $X$ cannot be simultaneous observers even if they are using separate identical sources and separate apparata! This cannot be true and we must look for another interpretation.
Suppose $Z$ and $X$ decide to combine their results and form a combined density matrix , $\rho^{(Z \emph{ with } X)}$ using (9) 
\begin{equation}
\rho^{(Z \emph{ with } X)}=(\frac{1}{2})(I + \langle \sigma_x \rangle \sigma_x + \langle \sigma_z \rangle \sigma_z )
\end{equation}
where $\langle \sigma_y \rangle=0$ in (9), by The Principle of Indifference $^9$ since neither $Z$ nor $X$ has any knowledge of the polarization in the $y$ direction. It is readily verified that the eigenvalues of $\rho^{(Z \emph{ with } X)}$ are $(1\pm\Gamma)/2$ where $\Gamma=+\sqrt[]{(\langle \sigma_x \rangle )^2 + (\langle \sigma_z \rangle )^2}$. We see that for these eigenvalues to be physically acceptable, then we must have $\Gamma \leq +1$ otherwise there must exist states with a negative occupation probability. Such probabilities violate the uncertainty principle ; the author has not managed to find a proof of this in any textbook, so a proof is given in Appendix D. So, it is impossible to find any physically realizable state for which $\Gamma > 1 $ . However, if we construct the density matrix using data, then violations of the uncertainty principle can occur. Suppose the observers $Z$ and $X$  collect data $D^{(Z)}=(n_{\uparrow}^{(Z)},n_{\downarrow}^{(Z)})$ and $D^{(X)}=(n_{\uparrow}^{(X)},n_{\downarrow}^{(X)})$ respectively. Using the same procedure as in section 2, the effective density matrix on combining the two data sets is
\begin{equation}
\rho_D^{(Z \emph{ with } X)}=\begin{pmatrix}
p^{(z,+1)} & \Delta_x \\
\Delta_x  & p^{(z,-1)}
\end{pmatrix}
\end{equation}
with
\begin{equation}
p^{(z, + 1)} = \frac{n_{\uparrow}^{(Z)} + 1 }{n^{(Z)} + 2 }
\end{equation}
\begin{equation}
p^{(z, - 1)} = \frac{n_{\downarrow}^{(Z)} + 1 }{n^{(Z)} + 2 }
\end{equation}
and
\begin{equation}
\Delta_x = \frac{1}{2}(p^{(x, + 1)}-p^{(x, - 1)})= \frac{1}{2}\frac{n_{\uparrow}^{(X)} - n_{\downarrow}^{(X)} }{n^{(X)} + 2 }
\end{equation}
Now it is readily verified that the eigenvalues of $\rho_D^{(Z \emph{ with } X)}$ lie outside the range $( 0,1)$ and , in particular , $\rho_D^{(Z \emph{ with } X)}$  has a negative eigenvalue , if
\begin{equation}
(\Delta_x)^2 \geq  p^{(z, + 1)}p^{(z, - 1)}
\end{equation}
and, when this inequality is satisfied, there is then a violation of the uncertainty principle. So if, for instance, the spin down detector used by observer $X$ is very inefficient, then the value for $(\Delta_x)^2$  will be close to $1/4$ which happens to be the maximum possible value of $p^{(z, + 1)}p^{(z, - 1)}$. So, if there is any significant polarisation of the $z$ component of the spin, (complete polarisation will give $ p^{(z, + 1)}p^{(z, - 1)}$ as zero) , then there will be a violation of the uncertainty principle caused by the inefficiency of $X$'s spin down detector.

On the other hand, in order to elucidate the Peierls commutation criterion,  suppose there is a third observer , $N$ , who, like $Z$ and $X$, has set up an identical source of silver atoms, and has put them through a Stern-Gerlach apparatus to split them into two beams, but then puts one of the beams through a second Stern-Gerlach apparatus with a much more powerful inhomogeneous magnetic field so that it splits into two further beams according to whether the nuclear spin is up or down. While this second apparatus is impractical, the nuclear magnetic moment is about 4 orders of magnitude smaller than the electron magnetic moment, it certainly illustrates the point Peierls makes. We suppose that observer $N$ wants to amalgamate their data with that of an observer $E$ who has measured the polarization of the electron spin of an identical source of silver atoms ; both $Z$ and $X$ are separately examples of observer $E$. Because the electron spin and the nuclear spin are commuting observables, we do not need to be specific about the directions of the inhomogeneous magnetic fields used by $E$ and $N$. We denote by $p_E^{(\pm 1)}$ the probabilities of electron spin up/down measured by $E$ and by $p_N^{(\pm 1)}$ the probabilities of nuclear spin up/down measured by $N$. The  density matrices for $E / N $ are given by
\begin{equation}
\rho_D^{(E/N)}= \begin{pmatrix}
p_{E/N}^{(+ 1)} & 0 \\
0 & p_{E/N}^{(- 1)}
\end{pmatrix}
\end{equation}
where the basis states are $(E/N,+1)$ corresponding to spin up and $(E/N,-1)$ corresponding to spin down.
Now consider how the two observers modify their density matrices in the light of the knowledge of the data of the other. To start with $E$ has no knowledge of $N$'s data and must allocate equal probabilities of $1/2$ for nuclear spin up/down. So in the basis of, in a self evident notation, $[(E,+),(N,+)]$ , $[(E,+),(N,-)]$ ,  $[(E,-),(N,+)]$ , $[(E,-),(N,-)]$ , they must have a density matrix
\begin{equation}
\rho_{D,0}^{(E)}= \begin{pmatrix}
p_{E}^{(+1)}/2 & 0              &0              &0               \\
0              & p_{E}^{(+1)}/2 &0              &0               \\
0              &0               &p_{E}^{(-1)}/2 &0               \\
0              &0               &0              &p_{E}^{(- 1)}/2
\end{pmatrix}
\end{equation}
while observer $N$ having no knowledge of the electron spin polarization has a density matrix
\begin{equation}
\rho_{D,0}^{(N)}= \begin{pmatrix}
p_{N}^{(+ 1)}/2 & 0                 & 0              & 0             \\
0               & p_{N}^{(-1)}/2    & 0              & 0             \\
0               &0                  &p_{N}^{(+ 1)}/2 &0              \\
0               &0                  &0               &p_{N}^{(-1)}/2
\end{pmatrix}
\end{equation}
These two matrices evidently commute as they are both diagonal. When each observer becomes aware of the other's data they update their respective density matrices to arrive at a common density matrix
\begin{equation}
\rho_D^{(both)}= \begin{pmatrix}
p_{E}^{(+1)}p_{N}^{(+1)} & 0                 & 0              & 0  \\
0                        &p_{E}^{(+1)}p_{N}^{(-1)}  & 0              & 0  \\
0                        &0                  &p_{E}^{(-1)}p_{N}^{(+1)} &0   \\
0                        &0             &0          &p_{E}^{(-1)}p_{N}^{(-1)}
\end{pmatrix}
\end{equation}
Now, in sharp contrast to the common density matrix arrived at by observers $Z$ and $X$ , this matrix will never violate the uncertainty principle no matter how inaccurate the data. The counters will always provide data such that the quantities $p_{E/N}^{(\pm 1)}$  lie between  zero and one and will obey $p_{E/N}^{(+ 1)} + p_{E/N}^{(- 1)} = 1$ so that $\Tr[\rho_D^{(both)}]=1$. The eigenvalues of $\rho_D^{(both)}$ will all be non-negative and no violation of the uncertainty principle is possible.

So the purpose of the Peierls commutation criterion is to ensure trouble free amalgamation of the data of two observers i.e. amalgamation without any danger of violation the uncertainty principle.

It is shown in appendix E that if the density matrices of two observers commute no matter how inaccurate the individual data, then the observables they are measuring must also commute, the same requirement from conventional quantum mechanics for two observables to be known precisely and simultaneously. This latter requirement is show to be necessary for imperfect measurements on a potentially mixed state system so that possible violations of the Uncertainty Principle can be avoided. In appendix F the immunity to uncertainty principle violations demonstrated above for electron and nuclear spin $1/2$ systems  is generalized to all systems with finite dimensional state spaces.

Peierls$^8$ also points out that the product of the density matrices of two observers should not be zero to avoid contradiction and the above analysis can illustrate this point. The product of the density matrices $\rho_{D,0}^{(E)}$ and  
 $\rho_{D,0}^{(N)}$ of the observers $E$ and $N$, is given by
 \begin{equation}
\rho_{D,0}^{(E)} \times \rho_{D,0}^{(N)}= \frac{1}{4}\begin{pmatrix}
p_{E}^{(+1)}p_{N}^{(+1)} & 0                 & 0              & 0  \\
0               & p_{E}^{(+1)}p_{N}^{(-1)}  & 0              & 0  \\
0              &0                  &p_{E}^{(-1)}p_{N}^{(+1)} &0   \\
0              &0                  &0             &p_{E}^{(-1)}p_{N}^{(-1)}
\end{pmatrix}
\end{equation}
If this product vanishes, each diagonal element must vanish and one finds that for that to be, at least , either both the probabilities $p_E^{(\pm 1)}$  or both the probabilities $p_N^{(\pm 1)}$ must vanish. This is clearly absurd and requiring that the product does not vanish is the minimum requirement to maintain coherence. The criterion that product of the density matrices of two observers should be non zero is discussed more generally in Appendix G.

\section{ Peierls Revisited}
In the light of the exposition in the previous sections, it is worthwhile revisiting Peierls's work, by quoting some of Peierls's final comments$^8$ and commenting on, and augmenting  them.

Firstly : 
\emph{“It is possible for two observers to have some knowledge of the same system, and the knowledge possessed by one may differ from the other. For example, one may only have an incomplete or inaccurate view of the measuring instrument. In this situation the two observers will use different density matrices.”}

This emphasises the subjective nature of the density matrix of an observer to the extent that the two observers are viewing the same measuring instrument rather than carrying out the same experiment in separate laboratories. Our treatment of the Stern Gerlach experiment in section 3 is directly applicable to either situation, each observer constructing their density matrix according to the counts they each actually record. It is clear that Peierls is only describing the very first stage of the scientific process here. With one observer having “only an incomplete or inaccurate view of the measuring instrument”, clearly the same and/or other observers will repeat the experiments : and provided only a small subset of the results are subject to systematic errors such as “an incomplete or inaccurate view of the measuring instrument”, their influence will be systematically squeezed out. The analysis in section 3 gives a simple working model to illustrate the process.

Secondly :
\emph{“However, the information possessed by the two observers must not contradict the uncertainty principle. For example, if one observer knows the z component of an atomic spin [exactly], the other may not know [ whether spin up or spin down is more likely for ] the x component. This is because the measurement made by the first observer would have caused an uncontrollable interference with the x component. This limitation can be expressed concisely by saying that the density matrices appropriate to the two observers must commute with each other.”}

The words inside the square brackets are the author's addition to aid clarity; and the $x$ and $z$ in the original have been interchanged for easier reference to the working in section 4 ( and, indeed, throughout this paper) in which the basis states have been the spin up and spin down states in the $z$ direction .

To visualize what Peierls is saying we imagine the first observer passing an atomic spin through a Stern Gerlach apparatus with an inhomogeneous magnetic field in the $z$ direction. If that first observer has a detector in both the spin up and spin down exit channels, then they will be able to find out the polarization of the $z$ component of the spin the atom had. But, it is no longer possible to do any other experiment on that atom. Suppose, instead of trying to record the spin, the first observer allows a second observer to set up a Stern Gerlach apparatus with inhomogeneous magnetic field in the $x$ direction  in the path along which the atom would emerge from the first observer's apparatus had it entered with spin up in the $z$ direction and for the second observer to set up detectors to measure the $x$ component of  the atomic spin; so we are now considering sequential measurements on the same system rather than parallel experiments on identical systems. Then according to section 4, because all counts corresponding to spin down in the $z$ direction are automatically excluded from the observations so that we know that the atom is entering the second observer's apparatus with spin up, we obtain a density matrix $\rho^{(Z \emph{ with } X)}$ of the form $
\begin{pmatrix}
1 &\Delta_x \\
\Delta_x  &0
\end{pmatrix}$ . Now this matrix will always have one negative eigenvalue and another greater than unity and hence is not an acceptable density matrix; it violates the uncertainty principle. The only exception is for $\Delta_x = 0 $ which corresponds to equal counts corresponding to spin up and spin down in the $x$ direction i.e. complete ignorance of the $x$ component of the spin. The precise knowledge of the $z$ component of the spin eliminates all possibility of knowing anything, except one's awareness of one's complete ignorance about the $x$ component.

If, however, the second observer measured the spin of the nucleus, then, according to our working in section 4, the two observers would arrive at the density matrix $
\begin{pmatrix}
p_{N}^{(+ 1)} & 0 \\
0 & p_{N}^{(- 1)}
\end{pmatrix}$ ( electron spin up assumed ) where $p_{N}^{(+ 1)}$ and $p_{N}^{(- 1)}$ are the probabilities deduced from the data for the $x$ component of the nuclear spin to be up and down respectively. As explained in section 4, these data dependent probabilities, can only lie between $0$ and 1 no matter how inaccurate the data and the matrix automatically satisfies the uncertainty principle, a property that is a consequence of the commutation of the electronic and nuclear spin variables. Only when Peierls's criterion is satisfied can one be sure that violations of the uncertainty principle will be avoided. It is not difficult to see that the root of the problem in the solely electronic measurements is the fact that $\sigma_x$ has only off diagonal elements and this is related to the fact that it does not commute with $\sigma_z$ .

Thirdly, Peierls continues to provide a justification of the commutation criterion :

\emph{“ This can be seen by considering the representation in which one of these [ density] matrices is diagonal. If the other  does not commute with it, it must have off-diagonal elements in the representation. This means that it shows phase relations between the states which the first observer could have observed, and this would violate the uncertainty principle.” 
}

We have given a fuller proof of this result in appendix H , but it is important to emphasize that it is necessary to assume that all the information in the density matrix has, in principle, subjective origins to establish the result. If one does not assume this, but assumes that the density matrix is derived from theory and not experimental data, then it is easy to find putative counter examples.

One can deduce from section 2 and  Appendices A and B  that the “... phase relations between the states ...” are none other than the expectation values of the observable measured by the second observer or by subsequent measurements by the first observer of that same observable. And because the measurements of the first and second observables are subject to errors there can be no guarantee that violations of the uncertainty principle will not occur.
And finally Peierls mentions his criterion to avoid a contradiction :

\emph{“ At the same time the two observers should not contradict each other. This means that the product of the two density matrices must  not be zero. 
Indeed, take a representation in which both are diagonal (which is possible if they commute); then there must be at least some states for which the probabilities assumed by both observers are non zero, and that means the product is non-zero. “
}

We have shown, in section 4, explicitly for two observers measuring respectively the electronic and nuclear spin of an atom that, if the product of the density matrices of the two observers vanishes, then we end up with an absurd situation : all probabilities for the measurement outcomes for at least one of the variables must vanish. And we have also expanded on the proof of Peierls's theorem in Appendix G. In particular, we point out that, if Peierls's criterion is not satisfied, it will be impossible to reconcile the two observers findings using further data and Bayes's Theorem.

\section{Summary}

We have provided an expansion and elucidation of Peierls's interpretation of quantum mechanics. Crucial to this has been the appreciation that the knowledge of an observer, encapsulated in their density matrix, can be expressed in terms of probabilities of observable outcomes even though those outcomes may be dependent on probability amplitudes and interference effects. As such, they can be updated as more data becomes available and eventually a consensus among the scientific community emerges.
In particular, it has been shown that the density matrix in any representation with a finite basis can be written in the terms of the expectation values of observable quantities and therefore in terms of probabilities of the outcome of the measurements being particular eigenvalues of those observables; the  case for two dimensions [ spin $1/2$ particles ] has been treated in detail and then generalized in appendices A and B. It has been shown how the density matrices can be constructed from experimental  data for the Stern Gerlach experiment and how two observers performing such experiments can collaborate and amalgamate their data both for the case in which they are measuring the same observable or two different observables and how large data sets can minimize the influence of initial prejudices, the essence of the scientific method. We have also shown that the criterion that the density matrices of two observers must commute is equivalent to the normal quantum mechanical condition that the two observables must commute if they are to be exactly known simultaneously; but the Peierls criterion shows that the relevance of the condition can be extended into the realm of individual imperfect, even wildly inaccurate measurements; if the two observables do not commute violations of the Uncertainty Principle may well occur . And finally , we have illustrated how the criterion that the product of the density matrices of two observers must be non zero is necessary for there to be any hope of reconciling the results of two observers with further experiment.

In finishing , the author would like to quote Peierls's statement$^8$ about his commutation criterion again, namely, 

\emph{“However, the information possessed by the two observers must not contradict the uncertainty principle. For example, if one observer knows the $z$ component of an atomic spin , the other may not know the $x$ component. This is because the measurement made by the first observer would have caused an uncontrollable interference with the $x$ component. This limitation can be expressed concisely by saying that the density matrices appropriate to the two observers must commute with each other.”}, 

a statement that has caused the author much puzzlement, and offer his own version, in which extra words are included in square brackets :

\emph{“However, the information possessed by the two observers must not contradict the uncertainty principle. For example, if one observer knows the $z$ component of an atomic spin [exactly], the other may not know [ whether spin up or spin down is more likely for  ] the $x$ component. This is because the measurement made by the first observer would have caused an uncontrollable interference with the $x$ component. This limitation can be [avoided and the condition necessary to do so can be ] expressed concisely by saying that the density matrices appropriate to the two observers must commute with each other [no matter how inaccurate their data].”
}

\section*{Appendix A : The Density Matrix in terms of the Expectation Values of Observables}

We assume that the state space has finite dimension, $N$. We will show how the real and imaginary parts of each and every element of the density matrix / operator , $\rho$ , can be expressed in terms of the expectation value , $ \langle O \rangle = \Tr[ \rho O]$ , of either some operator/ dynamical variable , $O$, or a linear combination of two such. There is a distinction between operators/ dynamical variables and observables in general$^{10}$ in that the latter must have a complete set of eigenvectors. Fortunately, since we are working in a state space of finite dimension, all dynamical variables are observables: hermitian operators are an example of normal operators (for which $AA^\dagger = A^\dagger A$) which are well known to have a complete set of orthonormal eigenvectors$^{11}$ even in the presence of degeneracy. The reader will readily be able to confirm that all the operators/dynamical variables used here are indeed observables. 

Introduce the $N$ observables $O(m,m)$ defined in some orthonormal basis (with vectors $|m \rangle , m = 1,2, \ldots ,N$ ) by

\begin{equation}
O(m,m) = |m \rangle \langle m |
\end{equation}
for which it is easily verified that for the diagonal components of $\rho$

\begin{equation}
\langle m |\rho |m \rangle  = \langle  O(m,m)  \rangle
\end{equation}

Moving on to expressing the off-diagonal elements of $\rho$ in terms of expectation values of observables, introduce the $N(N-1)/2$ observables ( $1 \leq k < l \leq N $)

\begin{equation}
O(k,l) = a|k \rangle \langle l | + a^*|l \rangle \langle k | 
\end{equation}
we readily find

\begin{equation}
\langle  O(k,l) \rangle  = a  \langle l|\rho| k\rangle  + a^* \langle k|\rho| l\rangle 
\end{equation}
If we now define observables $ O^{(x)}(k,l)$ and $ O^{(y)}(k,l)$ by setting $a = 1/2$ and
$a = -i/2$ in (42) respectively, then it is straightforward to find (for $1 \leq k < l \leq N $)

\begin{equation}
\langle k|\rho| l\rangle =  \langle  O^{(x)}(k,l) \rangle - i\langle  O^{(y)}(k,l) \rangle
\end{equation}
the elements $\langle l|\rho| k\rangle $ below the diagonal being just the complex conjugates.

\section*{Appendix B : Density Matrix in terms of  Probabilities of Observable Outcomes}

We have already demonstrated in Appendix A how the density matrix can be specified in terms of the expectation values of various observables.

As a straightforward consequence, one can also write the elements of the density matrix, $\rho$ , in an orthonormal basis ,  $|m \rangle $  $ (m = 1,2, \ldots ,N )$, in terms of the probabilities of various observational results alone.

For the diagonal elements, we have

\begin{equation}
\langle m |\rho |m \rangle  = \langle  O(m,m)  \rangle= p_m
\end{equation}
where $p_m$ is the probability of finding the system in the state $|m \rangle $

For off-diagonal elements in the upper triangle ( $ 1 \leq  k < l \leq N $)

\begin{equation}
\langle k|\rho| l\rangle =  \langle  O^{(x)}(k,l) \rangle - i\langle  O^{(y)}(k,l) \rangle
\end{equation}

Now the observables $ O^{(x)}(k,l)$ and $ O^{(y)}(k,l)$ each have two eigenstates each with non zero eigenvalue, $\sigma/2$ with $\sigma = \pm 1$  and $ N-2$ eigenstates each with zero eigenvalue. Obviously, in evaluating the expectation value of either observable the eigenstates corresponding to zero eigenvalue will not contribute and we have using $\alpha = (x,y)$  

\begin{equation}
\langle  O^{(\alpha)}(k,l) \rangle  = \sum_{\sigma} p^{\alpha,\sigma}_{k,l} \sigma /2
\end{equation}
or explicitly

\begin{equation}
\langle  O^{(\alpha)}(k,l) \rangle  = ( p^{\alpha,+1}_{k,l} -  p^{\alpha,-1}_{k,l}) /2
\end{equation}
where $p^{\alpha,\sigma}_{k,l}$  is the probability of the system being in the  eigenstate of $O^{(\alpha)}(k,l)$   with eigenvalue $\sigma /2$. Now introducing 

\begin{equation}
\Delta p^{\alpha}_{k,l}  = ( p^{\alpha,+1}_{k,l} -  p^{\alpha,-1}_{k,l})
\end{equation}

we can write for the elements above the diagonal,( $ 1 \leq k < l \leq N$ ) ,

\begin{equation}
\langle k|\rho| l\rangle = ( \Delta p^{(x)}_{k,l} - i \Delta p^{(y)}_{k,l}) /2
\end{equation}

And finally the complete density matrix may be written formally as

\begin{equation}
\rho = \sum_m  p_m O(m,m) + \sum_{k<l}  [\Delta p^{(x)}_{k,l} O^{(x)}(k,l) +  \Delta p^{(y)}_{k,l} O^{(y)}(k,l)]
\end{equation}

which is clearly expressed in terms of observables and probabilities of measurement outcomes.

\section*{ Appendix C : Bayes's Theorem and the Scientific Method}

We can use Bayes's theorem to evaluate the probability, $P[H\mid D]$ , of a scientific hypothesis, $H$ , being true, given that some data, $D$ , has been collected, and also given that the a priori probability that $H$ is true, is $P[H]$. The theorem states that 

\begin{equation}
P[H\mid D] = \frac{P[D\mid H]P[H]}{P[D]}
\end{equation}
where $P[D\mid H]$ is the probability that data , $D$ , would be collected if hypothesis , $H$ , were true. $P[D]$ is the probability that the data , $D$ , is collected regardless of whether $H$ is true or false :

\begin{equation}
P[D] = P[D\mid H]P[H] + P[D\mid \overline{H}]P[\overline{H}]
\end{equation}
where $\overline{H}$ denotes the opposite hypothesis to $H$ so that $P[H]+P[\overline{H}]=1$. One can now write Bayes's theorem as 

\begin{equation}
P[H\mid D] = \frac{1}{1+A}
\end{equation}
with 

\begin{equation}
A = \frac{P[D\mid \overline{H}]}{P[D\mid H]} \frac{P[\overline{H}]}{P[H]}
\end{equation}
Now suppose an experiment is set up to confirm the truth of hypothesis,$H$, such that 

\begin{equation}
P[D\mid H] >P[D\mid \overline{H}]
\end{equation}
If the experiment is repeated independently $n$ times, then the probability of this happening is $ (P[D\mid H])^n $ and $A$ has to be replaced by 

\begin{equation}
A_n = \Big(\frac{P[D\mid \overline{H}]}{P[D\mid H]}\Big)^n \frac{P[\overline{H}]}{P[H]}
\end{equation}
As $n$ becomes sufficiently large, $A_n $ tends to zero, and $P[H\mid D]$ tends inexorably towards unity regardless of the value of $P[H]$, i.e. the original prejudices of the observer(s)/scientific community. This simple result neatly encapsulates the scientific method. One notes that it is essential for $P[H]$ to differ from either zero or unity. If   $P[H]$ takes on either of these extreme values, then no amount of data collection will alter that belief.

\section*{Appendix D : Proof that a negative eigenvalue of the density matrix leads to a violation of the uncertainty principle.}

We first demonstrate the result for the spin $1/2$ case i.e. for 2 dimensions and then extend it to all finite dimensions. We work in a representation for which the density matrix is diagonal and write
\begin{equation}
\rho =  \begin{pmatrix}
\wp_1 &0\\
0 &\wp_2
\end{pmatrix}
\end{equation}
where $\wp_1$ and $\wp_2$ are the eigenvalues of the density matrix , $\rho$. Since one must have  $\Tr[\rho] = 1 $ , or $\wp_1+\wp_2=1$, at least one of $\wp_1$ and $\wp_2$ must be positive. We need to show that there exists at least one pair of observables , $A$  and $B$ , for which the uncertainty principle is violated i.e. for which

\begin{equation}
\langle(\Delta A)^2\rangle \langle(\Delta B)^2 \rangle < |\langle \Delta A  \Delta B \rangle|^2
\end{equation}
whenever one of $\wp_1$ and $\wp_2$ is negative. Here $\Delta O = O - \langle O \rangle I$ for any observable $O$. For notational convenience, take the $z$ axis so that the basis states that diagonalize $\rho $ correspond to spin up and spin down with respect to that axis and consider the difference of the left and right hand sides of the above inequality i.e. 

\begin{equation}
\langle(\Delta A)^2\rangle \langle(\Delta B)^2 \rangle - |\langle \Delta A  \Delta B \rangle|^2
\end{equation}
for the case $A = \sigma_x =\begin{pmatrix}
0 &1\\
1 &0
\end{pmatrix}$ and  $B = \sigma_y =\begin{pmatrix}
0 &-i\\
+i &0
\end{pmatrix}$ . It is readily verified that  $\langle \sigma_{\alpha} \rangle = \Tr[\rho \sigma_{\alpha} ]=0$ for $\alpha  =(x,y)$ so that $\Delta O = O $ for both $A$ and $B$ and hence
\begin{equation}
\langle (\Delta A)^2 \rangle = \langle ( \Delta \sigma_x )^2 \rangle= \langle (\sigma_x )^2 \rangle = \langle I \rangle = \wp_1 +\wp_2
\end{equation}
and similarly for $\langle (\Delta B)^2 \rangle $.
Therefore 
\begin{equation}
\langle(\Delta A)^2\rangle \langle(\Delta B)^2\rangle= (\wp_1 + \wp_2)^2
\end{equation}
On the other hand,
\begin{equation}
  |\langle \Delta A  \Delta B \rangle|^2 =  |\langle \sigma_x  \sigma_y  \rangle|^2 =  |\langle \sigma_z  \rangle|^2 = (\wp_1 - \wp_2)^2
\end{equation}
so that evaluation of $(60)$ gives
\begin{equation}
\langle(\Delta A)^2\rangle \langle(\Delta B)^2\rangle -  |\langle \Delta A  \Delta B \rangle|^2 = 4\wp_1\wp_2 < 0
\end{equation}
since $\wp_1$ and $\wp_2$ have opposite signs. So we have found a pair of observables for which the uncertainty principle is violated if $\rho$ has a negative definite eigenvalue. This is enough to establish the result that a negative eigenvalue for the density matrix can lead to a violation of the uncertainty principle for spin $1/2$ systems.

To extend the result to an arbitrary number of dimensions, $N$ , is straightforward. One notes that $\Tr[\rho]=1$ or $\wp_1+\wp_2 + \ldots + \wp_N = 1$ still requires at least one of the $\wp_i$'s to be positive. Relabel the eigenstates of $\rho$  so that $\wp_1$ is positive and $\wp_2$ negative. For the observables $A$ and $B$ use $ O^{(x)}(1,2)$ and $ O^{(y)}(1,2)$ introduced in appendix $A$ which are equal to $\sigma_x/2$ and $\sigma_y/2$ in the space spanned by the eigenvectors of $ \rho $ corresponding to the eigenvalues $\wp_1$ and $\wp_2$, but have all other elements zero. The proof of the violation of the uncertainty principle then proceeds as for the spin $1/2$ case.

\section*{Appendix E  : Peierls Commutation Criterion and Conventional Quantum Mechanics}

We demonstrate here that the Peierls Commutation Criterion, that the density matrices of two observers must commute for trouble free sharing of data, is the same as the conventional quantum mechanics requirement that two observables must commute if they are to be known precisely and simultaneously. The crucial assumption here is that the density matrices commute in all circumstances no matter what knowledge, inaccurate or even fanciful, each of the observers may have.
Now, the result would be a forgone conclusion if the density matrix could be an arbitrary function of the observable. In that case we could  just set $\rho^{(A)} =A$ and $\rho^{(B)} =B$ and the result follows. However, we have the constraints that $\rho$  is non-negative definite ( it must have non-negative eigenvalues) and that $\Tr[\rho]=1$. The subjective nature of  $\rho$  means that there are no other constraints. So suppose we consider

\begin{equation}
\rho^{(A)} = \frac{A+\lambda_A I}{\Tr[A+\lambda_A I]}
\end{equation}
where $\lambda_A$ is sufficiently large to make $\rho^{(A)}$ non-negative, i.e. larger than the modulus of any negative eigenvalue of $A$. We also have $\Tr[\rho^{(A)}]=1 $. With this assignment for $\rho^{(A)}$ and a similar one for $\rho^{(B)}$ , the condition $[\rho^{(A)},\rho^{(B)}] = 0 $ leads to 

\begin{equation*}
[\frac{A+\lambda_A I}{\Tr[A+\lambda_A I]},\frac{B+\lambda_B I}{\Tr[B+\lambda_B I]}]=0
\end{equation*}
or $[A,B] = 0 $ as in standard quantum mechanics. But rather than just ensuring that $A$ and $B$ may be known precisely and simultaneously, the condition $[A,B] = 0 $ is now seen to be necessary to avoid violations of the Uncertainty Principle when potentially inaccurate data sets from observers $A$ and $B$ are amalgamated.

\section*{Appendix F : Immunity from Uncertainty  Principle Violations }

We have demonstrated how an observer observing an electronic spin and a second observer observing a nuclear spin can combine their observations without fear of any violation of the uncertainty principle. We now want to demonstrate generally that two observers using commuting observables can do the same.

Since, when the density matrices, $\rho^{(A)}$ and $\rho^{(B)}$ , of two observers, A and B, commute, then their observables, $A$ and $B$, commute, we can construct a basis using the simultaneous eigenstates of $A$ and $B$. Observer A, in the absence of any knowledge of observer B's observations or knowledge will write the diagonal elements of their density matrix as 

\begin{equation}
p_{(a,b)}^{(A)}=p_{a}^{(A)} \times (\frac{1}{N_B})
\end{equation}
where $N_B$ is the number of eigenvalues of $B$ , since , while A from their own knowledge and observation can allocate a probability, $p_{a}^{(A)}$ , to a measurement of $A$ producing the result  $a$, they are ignorant of any measurements by B and must assign the same probability, $1/N_B$, to the result of any measurement of $B$. Similarly, observer B in the absence of any knowledge of observer A's observations or knowledge will write the diagonal elements of their density matrix as 

\begin{equation}
p_{(a,b)}^{(B)}=  (\frac{1}{N_A}) \times p_{b}^{(B)} 
\end{equation}
where $N_A$ is the number of eigenvalues of $A$ . If the two observers share their knowledge, they can each update their density matrix to 

\begin{equation}
\rho_{(a,b),(a',b')}^{(A \text{ }with\text{ } B)}=p_{a}^{(A)} \times p_{b}^{(B)} \delta_{(a,b),(a',b')}
\end{equation}
Because both $p_{a}^{(A)}$  and $p_{b}^{(B)} $ lie between zero and unity, this matrix can never have negative eigenvalues and hence the uncertainty principle cannot be violated even for wildly inaccurate probability assignments.

\section*{Appendix G : Peierls Product Criterion}

In section $4$ we demonstrated, for an example involving electronic and nuclear spins, how Peierls's product criterion must be obeyed to avoid absurd results. We now extend the discussion to higher dimensions.

The application of the product criterion assumes that the density matrices, $\rho^{(A)}$  and $\rho^{(B)} $ , of two observers, A and B, commute, so that we can use the result of the previous appendix, appendix F, to write the product of the two density matrices as
\begin{equation}
\rho_{(a,b)}^{(A)} \times \rho_{(a,b)}^{(B)}=\frac{ p_{a}^{(A)} \times p_{b}^{(B)}}{N_A N_B} 
\end{equation}
Peierls states$^{8}$ that the product of $\rho^{(A)}$ and $\rho^{(B)}$ cannot be zero in order to avoid contradiction and the above expression helps clarify this point. If the product of $\rho^{(A)}$ and $\rho^{(B)} $ is zero, then each and every product $ p_{a}^{(A)} p_{b}^{(B)}$  must be zero and consequently $\rho_{(a,b),(a',b')}^{(A \text{ }with\text{ } B)} =0$ as can be seen from the expression for $\rho^{(A \text{ }with\text{ } B)}$ given in appendix F. But this is impossible and, even worse, a situation which one cannot rectify by further experiment. To appreciate this, note that for the product $ p_{a}^{(A)} p_{b}^{(B)}$ to be zero at least one of $ p_{a}^{(A)}$  or $p_{b}^{(B)}$ must be zero. Suppose we assume $p_{a}^{(A)}\neq0$, then $p_{b}^{(B)}=0$  and, indeed, all $p_{b}^{(B)}=0$ and $\rho^{(B)} $ is just the null matrix, an absurd situation. The only course of action is to discard whatever knowledge led to B's probablility assignments of zero as they cannot be used as \emph{a priori} probabilities in any application of Bayes's theorem to update B's probablility assignments in the light of new data, as pointed out in Appendix C. So, the product of the density matrices of two observers must not be zero.

\section*{Appendix H : Violations of the Uncertainty Principle resulting from Non-commutivity of Density Matrices}

It was demonstrated in section 4 how violations of the uncertainty principle come about when using inaccurate data for a spin $1/2$ system. The cause of the violations was the combination of inaccurate data from two different observers measuring non-commuting observables. We will now demonstrate this result for a system in the general case i.e. with a state space of any finite dimension.

Let's call two observers A and B after the observables $A$ and $B$ they are individually measuring and denote by $\rho^{(A)}$ and $\rho^{(B)}$  their individual density matrices. In a basis using the eigenstates of $A$, $\rho^{(A)}$ is diagonal with elements $p_a^{(A)}$. The zeros in the off-diagonal elements represent A's ignorance and it is tempting for A to remove some of their ignorance by using part of B's data. Suppose A becomes aware that B assigns a value $\rho_{a,a'}^{(B)} \neq 0$ to the element $(a,a')$. Then A may well be tempted to incorporate that knowledge and modify their diagonal matrix to form

\begin{equation}
\begin{pmatrix}
\cdots &   \cdots               &\cdots              & \cdots & \cdots  \\
\cdots      & p_{a}^{(A)}  & \cdots              & \rho_{a,a'}^{(B)}  & \cdots\\
\cdots &   \cdots               &\cdots              & \cdots & \cdots  \\
\cdots &   \cdots               &\cdots              & \cdots & \cdots  \\
\cdots         &\rho_{a',a}^{(B)}   &\cdots       & p_{a'}^{(A)} &\cdots   \\
\cdots &   \cdots               &\cdots              & \cdots & \cdots  
\end{pmatrix}
\end{equation}
Now it is well known$^{12}$ that a matrix of this form must have 

\begin{equation}
p_{a}^{(A)} \times p_{a'}^{(A)} \geq |\rho_{a,a'}^{(B)} |^2
\end{equation}
if the eigenvalues are to all be non-negative and a violation of the uncertainty principle is to be avoided. But the problem is that the off-diagonal element $\rho_{a,a'}^{(B)}$, coming from B's observations, is independent of $p_{a}^{(A)} $  and $p_{a'}^{(A)} $ which come from A's observations, and there is no reason why the inequality should hold. Hence A's modification of their density matrix using B's data is in danger of violating the uncertainty principle.

\section*{Acknowledgements}
From the start of this work I have greatly benefited from detailed and illuminating discussions and correspondence with Drs D T Cornwell and B A Foreman. I am also indebted to Professors S J Clarke and T Lancaster for an important and enlightening discussion early on. I am also very grateful to Professors R B Stinchcombe and J T Chalker for their efforts , help and encouragement . And finally I would like to thank Professors S J Clarke and R A Abram for their general help and support with publication and helpful remarks on the manuscript.

\section*{References}

\begin{enumerate}
\item  Rudolf Peierls 1985  Bird of Passage  (Princeton University Press)
\item P.C.W. Davies and J.R. Brown  1986  (Eds.) The Ghost in the Atom  (Cambridge University Press)  pp 26 – 28
\item L D Landau and E M Lifshitz 1965 Quantum Mechanics (Permagon Press)  2rd Ed. § 1
\item R Peierls 1991 In defence of “measurement” Physics World  January pp 19–20
\item J S Bell  1990 Against “measurement”  Physics World  August pp 33 - 40
\item R Peierls 1979  Surprises in Theoretical Physics ( Princeton University Press) section 1.6
\item R Peierls 1985 Observations in Quantum Mechanics and the “ Collapse of the Wave Function” in Symposium on the Foundations of Modern Physics ( World Scientific)
\item R Peierls 1991  More Surprises in Theoretical Physics ( Princeton University Press) Section 1.2 especially pp 10 – 11.

\item E T Jaynes 2003 Probability Theory (Cambridge University Press) 
\item P A M Dirac 1958 The Principles of Quantum Mechanics 4th Edition (Oxford University Press) pp 26 \& 28
\item see e.g. M A Nielsen and I L Chang 2000 Quantum Computation and Quantum Information (Cambridge University Press) p 72

\item A Messiah 1999 Quantum Mechanics (Dover)  Problem 7 Chapter VII p293 

\end{enumerate}

\end{document}